\documentclass[letterpaper, 10 pt, conference]{ieeeconf} % Comment this line out
\IEEEoverridecommandlockouts                             % This command is only
\usepackage{amsmath} % assumes amsmath package installed
\usepackage{amssymb}  % assumes amsmath package installed

\newtheorem{mydef}{Definition}
\newtheorem{myrem}{Remark}

\usepackage{graphicx}
\usepackage{float}

\usepackage[linesnumbered,boxed,ruled,commentsnumbered]{algorithm2e}
\SetKwRepeat{Do}{do}{while}%
\usepackage{color}

\title{\LARGE \bf
Event-Triggered Distributed Model Predictive Control \\ \vspace{2pt} for Platoon Coordination at Hubs in a Transport System}

\author{Ting Bai,~Alexander Johansson,~Karl Henrik Johansson and Jonas Mårtensson% <-this % stops a space
\thanks{This work is supported by the Horizon 2020 through the project Ensemble, the Knut and Alice Wallenberg Foundation, the Swedish Foundation for Strategic Research and the Swedish Research Council.}\vspace{1.5pt}% <-this % stops a space

\thanks{T. Bai et al. are with the Integrated Transport Research Lab and Division of Decision and Control Systems, School of Electrical Engineering and Computer Science, KTH Royal Institute of Technology, Stockholm, Sweden, SE-100 44 Stockholm, Sweden. E-mails: \{{\tt\small tingbai, alexjoha, kallej, jonas1\}@kth.se}}}

\begin{document}

\maketitle
\thispagestyle{empty}
\pagestyle{empty}

%%%%%%%%%%%%%%%%%%%%%%%%%%%%%%%%%%%%%%%%%%%%%%%%%%%%%%%%%%%%%%%%%%%%%%%%%%%%%%%%
\begin{abstract}
This paper considers the problem of hub-based platoon coordination for a large-scale transport system, where trucks have individual utility functions to optimize. An event-triggered distributed model predictive control method is proposed to solve the optimal scheduling of waiting times at hubs for individual trucks. In this distributed framework, trucks are allowed to decide their waiting times independently and only limited information is shared between trucks. Both the predicted reward gained from platooning and the predicted cost for waiting at hubs are included in each truck's utility function. The performance of the coordination method is demonstrated in a simulation with one hundred trucks over the Swedish road network.
\end{abstract}

%%%%%%%%%%%%%%%%%%%%%%%%%%%%%%%%%%%%%%%%%%%%%%%%%%%%%%%%%
%-----------Section I. Introduction
\section{Introduction}
Truck platooning, which involves a group of trucks traveling in a formation with small inter-vehicle distances, has gained increasing popularity in modern transport systems \cite{janssen2015truck}. Seeing that every follower truck in a platoon experiences less aerodynamic drag than driving individually, platooning is promising for reducing trucks' fuel consumption, greenhouse gas emission and transport cost. These benefits have been recognized by numerous field experiments in, e.g., \cite{tsugawa2016review,al2010experimental}.

The technological advances in e-commerce developed over the past decade have made intelligent logistics a new application field for truck platooning. In this direction, efficient dispatching strategies for facilitating the formation of platoons while fulfilling delivery tasks attract more attentions. Specifically, researchers start to focus on the issues of planning the transport routes \cite{luo2018coordinated}, controlling the velocity of a truck that approaches to an intersection \cite{van2017efficient}, and scheduling the waiting times of a truck at hubs in transport networks \cite{larsen2019hub,rios2016survey} in order to maximize the benefit achieved from forming platoons. The decision makers of a scheduling process can be, for example, truck drivers, fleet managers, logistics companies, as well as administrators of the whole transport system.

In recent years, the ever-increasing demand for online shopping and economic globalization has led to significant growth of the amount of freight transport, especially for large-scale logistics companies with plenty of customers and trucks. Under this situation, traditional centralized dispatching methods become inapplicable. Apart from enduring a heavy computational burden, the centralized schemes suffer from a multi-fleet nature of the problem \cite{johansson2018multi}, where every truck has its own utility to optimize. Moreover, the centralized coordination approaches fail to efficiently deal with various delivery mission variations in the real-world application, for instance, adding new trucks to the system or changing the path planned or the shipping destinations. To handle these challenges, we aim to propose a novel distributed framework for solving the hub-based platoon coordination problem (i.e., platoons are formed at hubs) for large-scale transport systems.

%%%%%%%%%%%%%%%%%%%%%%%%%%%%%%%%%%%%%%%%%%%
%-----------Subsection A. Related Work
\subsection{Related Work}
The notion of truck platooning has been extensively investigated over the past few decades. In the literature, most research concentrates on the cooperative platooning control of vehicles from a physical layer, e.g., to form a platoon with arbitrary initial conditions \cite{tuchner2017vehicle}, maintain a constant speed and inter-vehicle space \cite{nowakowski2011cooperative}, enable multi-brand truck platooning~\cite{rodonyi2017adaptive}, apply vehicle-to-vehicle communication technique~\cite{jia2016platoon} and guarantee string stability \cite{besselink2017string,xiao2011practical}, etc. These research efforts primarily dedicated to maintaining a reliable running of truck platoons in the real traffic environment.

As platooning technology becoming mature, the development of efficient coordination strategies in the planning and scheduling layer has become a new active but challenging research topic \cite{bhoopalam2018planning,aarts2016european}. In this layer, existing studies involve exploring the frequent sub-routes in a network \cite{meisen2008data}, modeling the vehicle routing problem \cite{larsson2015vehicle} and planning the speed~\cite{van2015fuel} or the departure times~\cite{zhang2017freight,sokolov2017maximization} of vehicles to facilitate forming platoons. Notwithstanding, most existing methods are based on a centralized optimization and coordination mechanism, which therefore can not be applied to cope with massive networks with hundreds of vehicles. 

To make the time scheduling problem in large-scale systems computationally tractable, the literature~\cite{larson2014distributed} proposes to allocate virtual controllers at intersections and the platoon formation can be promoted by slightly adjusting the speed of vehicles approaching to intersections. This approach indeed decreases the computation load in comparison with centralized schemes, while as the platooning opportunities of individual trucks are only considered at one intersection, the platooning benefit at the following intersections may get lost. Another closely related work is~\cite{johansson2021strategic}, in which a similar waiting time scheduling problem is studied for trucks with individual utility functions. In difference to our work, the authors in~\cite{johansson2021strategic} formulate the problem in a non-cooperative game theoretic framework and present Nash equilibrium as solution concept, which requires a large number of iterations to find an equilibrium solution for many trucks. In this paper, we propose a solution that allows each truck to compute its waiting times independently from others, making it more efficient to compute than an equilibrium solution and more suitable for handling massive transport networks. 

%%%%%%%%%%%%%%%%%%%%%%%%%%%%%%%%%%%%%%%%%%%
%-----------Subsection B. Contributions
\subsection{Contributions}
In this paper, we aim to present a novel distributed framework for handling the truck platoon coordination problem in large-scale transport systems, where individual trucks are assumed to come from different commercial fleets and have their own utilities to optimize. The scheduling of trucks' waiting times at hubs is first formulated as an event-triggered distributed model predictive control (MPC) problem. Using this method, every truck optimizes its waiting times at hubs independently based on limited information from other trucks. The main contributions are as follows:
\begin{itemize}
    \item [(i)] A dynamical predictive model that characterizes trucks traveling in the transport network is created;
    \item [(ii)] A predicted utility function of individual trucks is presented, which includes the predicted reward obtained from platooning and the loss caused by waiting at hubs;
    \item [(iii)] An event-triggered distributed MPC algorithm is proposed to address the waiting time scheduling problem;
    \item [(iv)] The effectiveness of the developed method is verified through a platoon coordination simulation on one hundred trucks in the transport system of Sweden.
\end{itemize}
\vspace{1pt}

The rest of the paper is organized as follows. Section II formally formulates the platoon coordination problem and introduces basic terms used for transport network descriptions. Section III presents the main results of this paper, including a dynamical predictive model, the utility function design and an event-triggered distributed MPC algorithm. In Section IV, a simulation illustrating the effectiveness of the developed method is provided. Finally, Section V concludes this paper with an outlook for the future work.
\vspace{2pt}

%%%%%%%%%%%%%%%%%%%%%%%%%%%%%%%%%%%%%%%%%%%%%%%%%%%%%%%%%
%---------------Section II. Problem Formulation
\section{Problem Formulation}
\textit{Notations.} Throughout this paper, $\mathbb{R}$ and $\mathbb{N}$ denote the set of real numbers and non-negative integers, respectively. $\mathbb{R}^n$ is the $n$-dimensional Euclidean space while $\mathbb{R}^{n\times{m}}$ represents the set of $n\!\times\!{m}$ real matrices. A zero vector with appropriate dimension is denoted by $\textit{\textbf{0}}$. For any $a,b\!\in\!{\mathbb{N}}$ with $a\!\leq\!{b}$, let $[a\!:\!b]\!=\!\{a,a\!+\!1,\dots,b\}$. $|\mathcal{N}|$ stands for the cardinality of the set $\mathcal{N}$, i.e., the number of elements that in $\mathcal{N}$. 
   
\begin{figure}[thpb]
     \centering
     \includegraphics[width=0.7\linewidth]{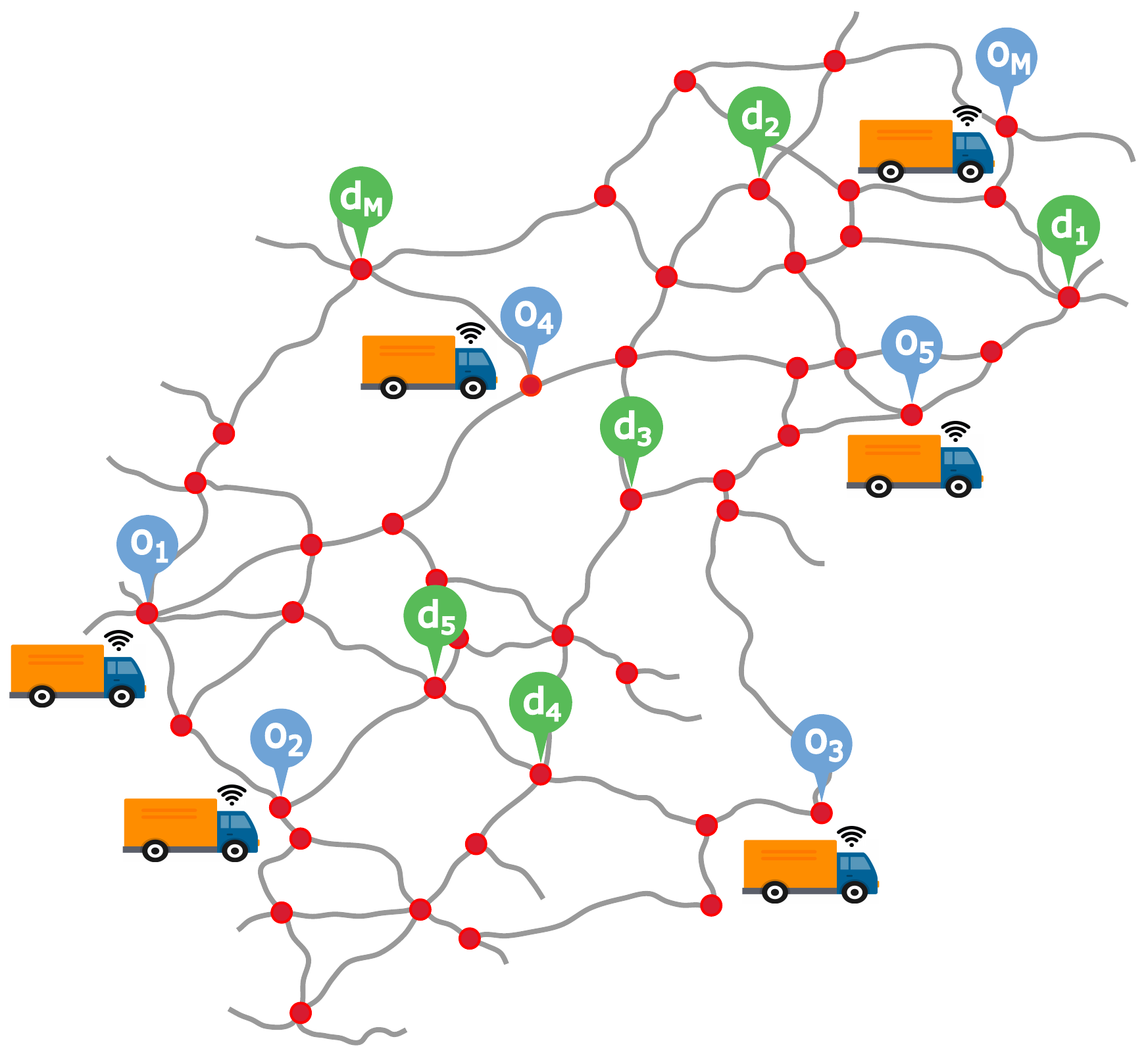}
     \vspace{-2pt}
      \caption{A graph illustrating the platoon coordination problem considered in this paper, where hubs in the transport network are denoted by red nodes. Origins and destinations are marked by blue and green labels, respectively.} \vspace{-2pt}
      \label{Fig.1}
   \end{figure}

As shown in Fig.~\ref{Fig.1}, in this paper, we focus on a large-scale transport system with $M$ trucks from different fleets. Every truck travels from its origin $o_i$ to its destination $d_i$ to achieve a certain delivery task, where $i\!\in\!{\mathcal{M}:=[1\!:\!M]}$. 
\vspace{1pt}
   
We make use of some assumptions to simplify the presentation of the problem. For every truck $i$, its delivery route from the origin $o_i$ to the destination $d_i$, the hubs alongside the route, as well as the travel time on routes are assumed to be fixed and known. The formation of platoons takes place only at hubs. On this basis, we formulate \textbf{the platoon coordination problem at hubs} as: optimize the waiting times of every individual truck $i$ at hubs alongside its delivery route such that its utility is maximized. 
   
\begin{myrem}
 For simplicity, we assume all the origin and destination pairs (namely, OD pairs) are selected from hubs. In other words, every origin (and destination) in the transport network corresponds to a hub in our consideration.
\end{myrem}
\vspace{3pt}

In the sequel, we present necessary definitions that will be used in describing the transport network. We consider that the route of truck $i$ is partitioned by a group of hubs $\mathcal{H}_i\!=\![0\!:\!N_i\!-\!1]$ into a finite number of road segments. The origin $o_i$ is indexed by $0$-th hub while the destination $d_i$ is indexed by $(N_i\!-\!1)$-th hub. For any hub $k\!\in\![0,N_i\!-\!2]$ in the route of truck $i$, we denote by $\textit{e}_{i(k,k+1)}$ the \textit{directed route segment} of truck $i$ from hub $k$ to the next hub $k\!+\!1$. The \textit{route} of truck $i$, denoted by $\textit{\textbf{e}}_{(o_i,d_i)}$, is defined as a sequence of directed route segments from $o_i$ to $d_i$, that is,
\begin{equation}
    \textit{\textbf{e}}_{(o_i,d_i)}:=\{\textit{e}_{i(0,1)},\textit{e}_{i(1,2)},\dots,\textit{e}_{i(N_i-2,N_i-1)}\}.\label{equ.1}
\end{equation}

Meanwhile, we associate the global transport system with a directed graph $\mathcal{D}(\mathcal{H},\mathcal{E})$ for route structure analysis, where $\mathcal{H}$ is the set of hubs while $\mathcal{E}$ is the set of directed route segments in the network. The total number of hubs is $|\mathcal{H}|\!=\!N$. Every physical hub in the transport system is represented by $h_m$ with $m\!\in\![1\!:\!N]$. On this basis, $\mathcal{H}$ and $\mathcal{E}$ are denoted as
\begin{equation}
    \mathcal{H}=\{h_1,h_2,\dots,h_N\},\quad \mathcal{E}=\cup_{i=1}^M\textit{\textbf{e}}_{(o_i,d_i)}.\label{equ.2}
\end{equation}

Notice that the same hub index $k$ in the routes of different trucks may correspond to different real physical hubs. Here we do not distinguish the hub indexes for brevity. To avoid confusion, we make use of a map $H_i(k)$ to represent the corresponding physical hub. More specifically, the function $H_i$:~$\mathcal{H}_i\!\to\!{{\mathcal{H}}}$ maps any hub index $k\!\in\!\mathcal{H}_i$ in the route of truck $i$ into a real physical hub $h_m\!\in\!{\mathcal{H}}$ in the transport network. 

In line with the above notions, the \textit{common route segment} of two routes can be presented as follows. For any two route segments $\textit{e}_{i(k,k+1)}\!\in\!{\textit{\textbf{e}}_{(o_i,d_i)}}$ and $\textit{e}_{j(k^{'},k^{'}+1)}\!\in\!{\textit{\textbf{e}}_{(o_j,d_j)}}$ where $i,j\!\in\!{\mathcal{M}}$ and $i\!\neq\!{j}$, we say $\textit{e}_{i(k,k+1)}\!=\!\textit{e}_{j(k^{'},k^{'}\!+1)}$ is a common route segment for the routes $\textit{\textbf{e}}_{(o_i,d_i)}$ and $\textit{\textbf{e}}_{(o_j,d_j)}$ if $H_i(k)\!=\!H_j(k^{'})$ and $H_i(k\!+\!1)\!=\!H_j(k^{'}\!\!+\!1)$.
\vspace{3pt}

%%%%%%%%%%%%%%%%%%%%%%%%%%%%%%%%%%%%%%%%%%%%%%%%%%%%%%%%%
%-------------------Section III. Distributed MPC for Platoon Coordination
\section{Event-triggered Distributed MPC for Platoon Coordination}
This section introduces how one can solve the platoon coordination problem at hubs through an event-triggered distributed MPC approach. To this end, we start by presenting the dynamical predictive model of individual trucks.

%%%%%%%%%%%%%%%%%%%%%%%%%%%%%%%%%%%%%%%%%%%%%%%%%%%%%%%%%%%%%%
%----------Subsection A. Dynamical Model for Individual Trucks
\subsection{Dynamical Model of Individual Trucks}
The distributed MPC has been regarded as an advanced control method for optimal control of networked systems due to its capability of handling multiple constraints and improving system performance with limited local information~\cite{farina2012distributed,bai2020distributed,dunbar2007distributed}. To employ the distributed MPC algorithm, a dynamical subsystem model for prediction is required.
\vspace{0.5pt}

\begin{figure}[thpb]
     \centering
     \includegraphics[width=0.999\linewidth]{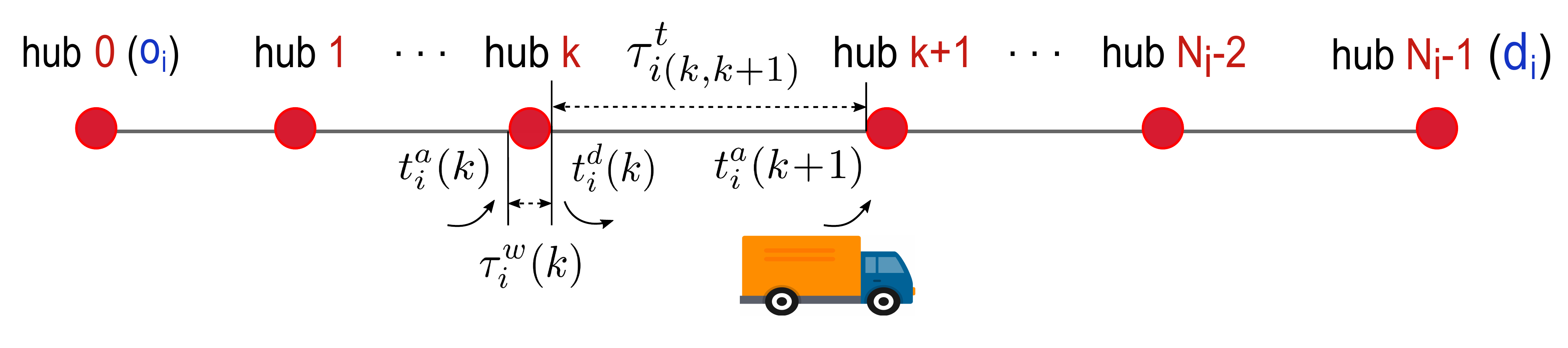}
     \vspace{-2pt}
      \caption{The route of truck $i$ from the origin $o_i$ to the destination $d_i$.}
      \label{Fig.2}
   \end{figure}
 
Let us consider a large-scale distributed transport system that consists of $M$ trucks, where each truck is regarded as a local subsystem. For every truck $i\!\in\!{\mathcal{M}}$, its departure time $t_i^d(k)$ from hub $k\!\in\![0\!:\!N_i\!-\!2]$ is denoted as
\begin{equation}
    t_i^d(k)=t_i^a(k)+\tau_i^w(k),\label{equ.3}
\end{equation}
where $t_i^a(k)$ represents the arrival time of truck $i$ at hub $k$ and $\tau_i^w(k)$ denotes its waiting time at hub $k$. As shown in Fig.~\ref{Fig.2}, the arrival time of truck $i$ at the next hub $k\!+\!1$ can be depicted by
\begin{equation}
    t_i^a(k\!+\!1)=t_i^d(k)+\tau_{i(k,k+1)}^t,\label{equ.4}
\end{equation}
where $\tau_{i(k,k+1)}^t$ is the travel time of truck $i$ on the route segment $\textit{e}_{i(k,k+1)}$. By substituting equation (\ref{equ.3}) in equation (\ref{equ.4}), we have that
\begin{equation}
    t_i^a(k\!+\!1)=t_i^a(k)+\tau_i^w(k)+\tau_{i(k,k+1)}^t,~k\!\in\![0\!:\!N_i\!-\!2].\label{equ.5}
    \end{equation}

In (\ref{equ.5}), we take $t_i^a$ and $\tau_i^w$ as the state $x_i$ and the input $u_i$ for truck $i$, respectively. Because the travel time $\tau_{i(k,k+1)}^t$ is known in the problem setting, it can be denoted as a constant $c_i$. Therefore, the dynamical model of truck $i$ is of the form
\begin{equation}
    x_i(k\!+\!1)=x_i(k)+u_i(k)+c_{i(k,k+1)},~ k\!\in\![0\!:\!N_i\!-\!2],\label{equ.6}
\end{equation}
where $x_i, u_i\!\in\!{\mathbb{R}}$ are the arrival and waiting times of truck $i$. Constrained by the delivery deadline, the input of every subsystem, namely, the waiting time of a truck at each hub, should satisfy that $u_i\!\in{\mathcal{U}_i}:={\{u_i:\underline{\tau}_i^w\leq{u_i}\leq{\bar{\tau}_i^w}\}}$. 
\vspace{2pt}

The dynamical model (\ref{equ.6}) characterizes the dynamics of every truck traveling in the transport network, which indicates that one can predict the arrival time of a truck at the next hub in accordance with the arrival and waiting times of the truck at the current hub, as well as the travel time on the route segment between the two hubs. 

%%%%%%%%%%%%%%%%%%%%%%%%%%%%%%%%%%%%%%%%%%%%%%%%%%%%%
%-----Subsection B. Utility Function
\subsection{Utility Function}
In a transport network system, trucks are assigned to deliver goods from their own origins to destinations within a preferred time range. We assume that each truck from a different commercial fleet and thus has a different utility function to optimize. The target of the platoon coordination is to schedule the waiting times at hubs optimally for each individual truck so that its own utility is maximized.
\vspace{2pt}

Typically, the utility function of a truck contains the reward gained from forming platoons with other trucks and the loss caused by waiting at hubs. According to the dynamical model of individual trucks obtained above, the dynamics of a truck at the future hubs become predictable. Therefore, the \textit{predicted reward function} and the \textit{predicted loss function} of every truck are presented as follows.
\vspace{5pt}
\begin{itemize}
\item \textit{The Predicted Reward Function:}
\end{itemize}
\vspace{4pt}

%------------------The Predicted Reward Function
The truck platooning technique is essentially a cooperative behavior among a group of trucks, which requires all trucks in the platoon to: \textit{(c1)} start from a common hub and leave for the next common hub; and \textit{(c2)} have the same departure time from the common hub. To denote all the trucks in the transport system that can meet condition (c1), the \textit{potential partner set} of a truck is defined.
\vspace{3pt}
\begin{mydef}
(\textit{Potential partner set}) 
For two trucks $i,j\!\in\!{\mathcal{M}}$ and $i\!\neq\!{j}$, given $\textit{e}_{i(k,k+1)}\!\in\!{\textit{\textbf{e}}_{(o_i,d_i)}}$ with $k\!\in\![0\!:\!N_i\!-\!2]$, if $\textit{e}_{i(k,k+1)}\!\in\!{\textit{\textbf{e}}_{(o_j,d_j)}}$ holds, we say truck $j$ is a potential platooning partner of truck $i$ at hub $k$. The potential partner set of truck $i$ at hub $k$ is denoted as
\begin{equation}
    \mathcal{P}_i(k)\!=\!\{j\!:\textit{e}_{i(k,k+1)}\!\in\!{{\textit{\textbf{e}}_{(o_j,d_j)}}},~j\!\in\!{\mathcal{M}},j\!\neq\!{i}\}.\label{equ.7}
\end{equation}
\end{mydef}
\vspace{5pt}

If truck $j$ belongs to the potential partner set of truck $i$, truck $j$ is also called a \textit{related truck} of truck $i$. Here $\mathcal{P}_i(k)$ represents all the trucks that may form a platoon with truck $i$ on the common route segment $\textit{e}_{i(k,k+1)}$.
\vspace{2pt}

\begin{myrem}
The potential partner set $\mathcal{P}_i(k)$ can be calculated offline based on the routes information of trucks. If a truck $j$ in $\mathcal{P}_i(k)$, it implies that truck $i$ will consider truck $j$ as a potential platooning partner at its $k$-th hub since they have the route segment $\textit{e}_{i(k,k+1)}$ in common.
\end{myrem}
\vspace{2pt}

Next, to meet the condition (c2) for forming platoons, the \textit{predicted partner set} of every truck is given.
\vspace{3pt}
\begin{mydef}
(\textit{Predicted partner set}) For any hub $k\!\in\![0\!:\!N_i\!-\!2]$ in the route $\textbf{\textit{e}}_{(o_i,d_i)}$, the predicted partner set of truck $i$ at the future hub $k\!+\!h$ in $\textbf{\textit{e}}_{(o_i,d_i)}$ is denoted by
\begin{align}
    \mathcal{R}_i(k\!+\!h|k)\!=\!\{&j:j\!\in\!{\mathcal{P}_i(k\!+\!h)}\wedge x_i(k\!+\!h|k)\!+\!u_i(k\!+\!h|k)\nonumber\\
    &\quad =\hat{x}_j(H_i(k\!+\!h))\!+\!\hat{u}_j(H_i(k\!+\!h))\},\label{equ.8}
\end{align}
with $h\!\in\![0\!:\!N_i\!-\!2\!-\!k]$. In (\ref{equ.8}), $x_i(k\!+\!h|k)$ and $u_i(k\!+\!h|k)$ denote the state and input of truck $i$ at the future hub $k\!+\!h$ predicted by truck $i$ at the current hub $k$, respectively. Whereas $\hat{x}_j(H_i(k\!+\!h))$ and $\hat{u}_j(H_i(k\!+\!h))$ denote the state and input of truck $i$'s related trucks $j$ at the physical hub $H_i(k\!+\!h)$ predicted by truck $j$.
\end{mydef}
\vspace{2pt}

\begin{figure}[thpb]
     \centering
     \includegraphics[width=1\linewidth]{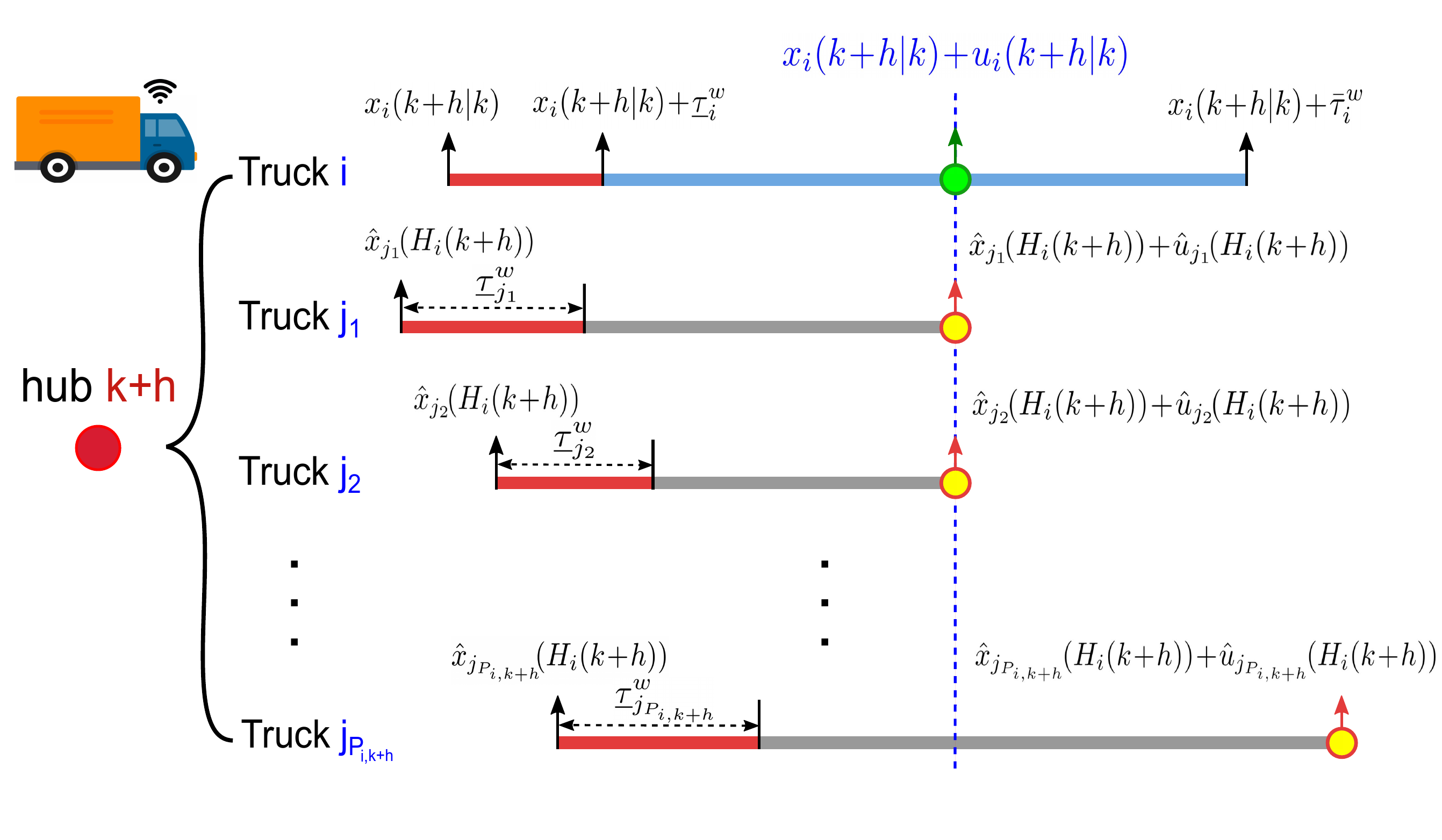}
     \vspace{-20pt}
      \caption{The waiting time bars of truck $i$ and $j$ at hub $k\!+\!h$ in the route $\textbf{\textit{e}}_{(o_i,d_i)}$, where $j\!\in\!{\mathcal{P}_i(k\!+\!h)}$ and $|\mathcal{P}_i(k\!+\!h)|\!=\!P_{i,k+h}$.}
      \label{Fig.3}
      \vspace{-5pt}
   \end{figure}

Regarding to the definition of the predicted partner set $\mathcal{R}_i(k\!+\!h|k)$, we explain the following things. For clarity, Fig.~\ref{Fig.3} is used in the explanation.

\begin{itemize}
  \item[1).] In (\ref{equ.8}), $x_i(k\!+\!h|k)\!+\!u_i(k\!+\!h|k)$ represents the departure time of truck $i$ from the hub $k\!+\!h$ predicted at hub $k$. Its feasible range is marked by the blue time bar in Fig.~\ref{Fig.3}. Similarly, $\hat{x}_j(H_i(k\!+\!h))+\hat{u}_j(H_i(k\!+\!h))$ denotes the departure time of truck $i$'s related trucks (as shown by the yellow nodes in Fig.~\ref{Fig.3}) from the real physical hub $H_i(k\!+\!h)$. Therefore, the equation in (\ref{equ.8}) indicates that truck $i$ is able to form a platoon with all the trucks $j\!\in\!{\mathcal{R}_i(k\!+\!h|k)}$ at the hub $k\!+\!h$ in $\textbf{\textit{e}}_{(o_i,d_i)}$ according to the prediction of truck $i$ at the current hub $k$;
  \vspace{5pt}
 
  \item[2).] The hub index $k\!+\!h$ in the route $\textbf{\textit{e}}_{(o_i,d_i)}$ of truck $i$ may differ from the hub index in the route $\textbf{\textit{e}}_{(o_j,d_j)}$ of the related trucks $j$. To denote the prediction of truck $j$ at the same hub $k\!+\!h$ in $\textbf{\textit{e}}_{(o_i,d_i)}$, the function $H_i(k\!+\!h)$ is used to map the hub $k\!+\!h$ to a real physical hub in the transport network in (\ref{equ.8}). Meanwhile, at which hub in $\textbf{\textit{e}}_{(o_j,d_j)}$ the state and input $\hat{x}_j(H_i(k\!+\!h))$, $\hat{u}_j(H_i(k\!+\!h))$ are predicted by truck $j$ is not important for truck $i$ and thereby is not included in the notations.
\end{itemize}
\vspace{5pt}

Extensive field experiments \cite{lu2011automated,davila2013environmental,bishop2017evaluation} have shown that the follower trucks in a platoon have approximately the same fuel savings, while the leader truck gets a significantly smaller saving. The total benefit of a platoon is therefore increasing with the number of followers and the platoon's duration. For this reason, the predicted reward function of an individual truck $i$ at its $k$-hub is defined by   
\vspace{1pt}
\begin{equation}
    R_i(k):=\sum_{h=0}^{N_i-2-k}\xi_ic_{i(k+h, k+h+1)}\frac{|\mathcal{R}_i(k\!+\!h|k)|}{|\mathcal{R}_i(k\!+\!h|k)|\!+\!1},\label{equ.9}
\end{equation}
which includes the rewards achieved for forming platoons on the remaining route segments of truck $i$, that is, $\{\textit{e}_{i(k,k+1)},\textit{e}_{i(k+1,k+2)},\dots,\textit{e}_{i(N_i\!-\!2,N_i\!-\!1)}\}$. More specifically, $\xi_i$ represents the monetary platooning benefit from fuel savings per follower truck and per travel time unit, and $c_{i(k+h,k+h+1)}$ is the travel time on the route segment $\textit{e}_{i(k+h,k+h+1)}$. The cardinality of $\mathcal{R}_i(k\!+\!h|k)$ is the predicted number of follower trucks in the platoon forming at hub $k\!+\!h$. For simplicity, we assume that the leader has zero fuel saving and the total platooning benefit is shared equally in the $|\mathcal{R}_i(k\!+\!h|k)|\!+\!1$ trucks in the platoon.
\vspace{3pt}

\begin{myrem}
Unlike $\mathcal{P}_i(k)$, the predicted partner set $\mathcal{R}_i(k\!+\!h|k)$ includes the set of trucks that truck $i$ predicts to form a platoon at hub $k\!+\!h$ according to the predicted arrival times and waiting times of other related trucks. It is worth mentioning that $\mathcal{R}_i(k\!+\!h|k)$ is a subset of $\mathcal{P}_i(k\!+\!h)$.

\end{myrem}
\vspace{4pt}
\begin{itemize}
    \item \textit{The Predicted Loss Function:}
\end{itemize}
\vspace{3pt}

%----------The Predicted Loss Function
The trucks can wait at hubs to form platoons and their decisions to wait may incur benefit loss because of higher cost for drivers and the penalties for delay in the transport operation. Taking into account these factors, we define the predicted loss function for truck $i$ at hub $k$ as
\begin{equation}
    L_i(k):=\sum_{h=0}^{N_i-2-k}\epsilon_iu_i(k\!+\!h|k),\label{equ.10}
\end{equation}
where $\epsilon_i$ denotes the monetary loss per time for waiting and $u_i(k\!+\!h|k)$ denotes the waiting time of truck $i$ at hub $k\!+\!h$ computed at hub $k\!\in\![0:N_i\!-\!2]$.

Then, the utility function $J_i(k)$ of truck $i$ at the $k$-th hub in its route is comprised of its predicted reward function and loss function, that is
\begin{align}
    J_i(k)&=R_i(k)-L_i(k)\label{equ.11}\\
          &=\!\!\sum_{h=0}^{N_i-2-k}l_i\big
          (x_i(k\!+\!h|k), u_i(k\!+\!h|k), \nonumber\\
          &\quad\quad\quad\quad\quad~ \hat{\textit{\textbf{x}}}_{-i}(H_i(k\!+\!h)), \hat{\textit{\textbf{u}}}_{-i}(H_i(k\!+\!h))\big
),\label{equ.12}
\end{align}
where $\hat{\textit{\textbf{x}}}_{-i}(H_i(k\!+\!h))$ and $\hat{\textit{\textbf{u}}}_{-i}(H_i(k\!+\!h))$ are the collection of predicted states and inputs of potential platooning partners of truck $i$. More precisely, $\hat{\textit{\textbf{x}}}_{-i}$ and $\hat{\textit{\textbf{u}}}_{-i}$ are denoted by
\vspace{-3pt}
\begin{align}
    &\hat{\textit{\textbf{x}}}_{-i}(\!H_i(k\!+\!h))\!=\![\hat{x}_j(H_i(k\!+\!h))]_{\forall{j}\in{\mathcal{P}_i(k+h)}}\nonumber\\
    &\quad\quad\quad\quad\quad\quad\!=\![\hat{x}_1\!(H_i(k\!+\!h)),\!...,\!\hat{x}_{P_{i,k+h}}\!(H_i(k\!+\!h))]\!\!\!\label{equ.13}\\
    &\hat{\textit{\textbf{u}}}_{-i}(\!H_i(k\!+\!h))\!=\![\hat{u}_j(H_i(k\!+\!h))]_{\forall{j}\in{\mathcal{P}_i(k+h)}}\nonumber\\ &\quad\quad\quad\quad\quad\quad\!=\![\hat{u}_1\!(H_i(k\!+\!h)),\!...,\!\hat{u}_{P_{i,k+h}}\!(H_i(k\!+\!h))]\!\!\!\label{equ.14}
\end{align}
with $P_{i,k+h}\!=\!|\mathcal{P}_i(k\!+\!h)|$.
\vspace{2pt}

%%%%%%%%%%%%%%%%%%%%%%%%%%%%%%%%%%%%%%%%%%%%%%%%%%%%%%
%------------Subsection C. Event-triggered Distributed MPC for Platooning
\subsection{Event-triggered Distributed MPC for Platooning}
\vspace{1pt}

In the hub-based platoon coordination problem considered in this paper, every individual truck $i\!\in\!{\mathcal{M}}$ in the transport network is allowed to make decisions only when it arrives at hubs. The decision variables are the waiting times of truck $i$ at hubs, and the optimization target is its own utility function. In this sense, the distributed platoon coordination problem at hubs is in nature an event-triggered control problem. The \textit{Event-triggering Condition} can be depicted as
\begin{align}
    t_{sys}\!=\!x_i(k)~\text{and}~k\!\neq\!{N_i\!-\!1},~~ i\!\in\!{\mathcal{M}},\label{equ.15}
\end{align}
where $t_{sys}$ is the system time of the whole transport system and $x_i(k)$ is the arrival time of truck $i$ at any a hub $k$ except for destination. That is, $k\!\neq\!{N_i\!-\!1}$. 

On the basis of the above descriptions and definitions, the optimization problem of truck $i$ at hub $k\!\in\![0:N_i\!-\!2]$ in its route can be formulated as the following distributed MPC problem
\begin{subequations}
\begin{align}
    \max_{\textit{\textbf{u}}_i(k)}~\quad  &J_i(k)\label{equ.16a}\\
    \mathrm{s.t.}~\quad &x_i(k)=t_i^a(k)\label{equ.16b}\\
    &x_i(k\!+\!1|k)=x_i(k)+u_i(k)+c_{i(k,k+1)}\label{equ.16c}\\
    &u_i(k\!+\!h|k)\in{\mathcal{U}_i},~~h\!\in\![0\!:\!N_i\!-\!2\!-\!k]\label{equ.16d}\\
    &x_i(N_i\!-\!1|k)-t_i^{end}\!\leq{0}.\label{equ.16e}
\end{align}
\end{subequations}

In problem (16), the utility function $J_i(k)$ is given in (\ref{equ.12}), which is dependent on the departure time predictions of truck $i$'s related trucks. The optimization variable $\textit{\textbf{u}}_i(k)\!\in\!{\mathbb{R}^{N_i\!-1\!-k}}$ has the specific form of
\begin{equation}
    \textit{\textbf{u}}_i(k):=[u_i(k),u_i(k\!+\!1|k),\dots,u_i(N_i\!-\!2|k)].\label{equ.17}
\end{equation}
We denote the optimal solution of problem (16) as $\textit{\textbf{u}}^{*}_i(k)\!=\![u_i^{*}(k),\dots,u_i^{*}(N_i\!-\!2|k)]$, where the elements are the optimal waiting times at the hubs in truck $i$'s route. It is important to point out that only the optimal waiting time $u_i^{*}(k)$ computed for the current hub $k$ will be implemented in the control of truck $i$. The other waiting times will serve as predictions for other related trucks to compute their optimal waiting times.
\vspace{3pt}

The constraints (\ref{equ.16b})-(\ref{equ.16e}) are explained as follows: (\ref{equ.16b}) represents the initial state of truck $i$ at the current hub. Particularly, there is $t_i^a(k)\!=\!t_i^{start}$ when $k\!=\!0$, where $t_i^{start}$ denotes the departure time of truck $i$ from its origin. In addition, (\ref{equ.16c}) is the dynamical predictive model of truck $i$ and (\ref{equ.16d}) is the restrictions on the waiting times. Finally, (\ref{equ.16e}) implies that every truck respects its delivery deadline at its destination. The developed distributed MPC method for platoon coordination at hubs is organized as in \textbf{Algorithm}~\textbf{\ref{Alg.1}}, where $\mathcal{S}_d$ denotes the set of decision makers at time $t_{sys}$.

\IncMargin{0.57em}  
\begin{algorithm}
\SetKwInOut{Input}{\textbf{Input}}\SetKwInOut{Output}{\textbf{Output}} % 替换关键词
    \Input{$\mathcal{D}(\mathcal{H},\mathcal{E})$, $t_i^{start}$, $t_i^{end}$, $\tau_{i(k,k+1)}^t$, $\underline{\tau}_i^w$, $\bar{\tau}_i^w$}
    \Output{ $\textbf{\textit{u}}_i^*(k)$, $\{\mathcal{R}_i^*(k\!+\!h|k)\}$}
    \BlankLine
   {Initialization: $\textit{\textbf{u}}_i^*(0)\leftarrow \textbf{\textit{0}}$, obtain
   $x_i(k)$, $\hat{\textit{\textbf{x}}}_{-i}(H_i(k))$, $\hat{\textit{\textbf{u}}}_{-i}(H_i(k))$}, $\mathcal{P}_i(k)$\; 
   $\mathcal{S}_d\leftarrow{\emptyset}$\;
   $t_{sys}\leftarrow{0}$\;\vspace{1pt}
  \While{$t_{sys}\neq{\max_{i\in{\mathcal{M}}}{\{t_i^{end}\}}}$}{  
  \vspace{2pt}
  $t_{sys}\leftarrow{t_{sys}\!+\!1}$\;
  \vspace{2pt}
  $\mathcal{S}_d\leftarrow{\{i\!\in\!{\mathcal{M}: t_{sys}\!=\!x_i(k) \wedge k\!\neq\!{N_i\!-\!1}}\}}$\;
  \vspace{3pt}
  
   \For{$i\!\in\!{\mathcal{S}_d}$}{
   \vspace{2pt}
    \!solve truck $i$'s distributed MPC problem (16)\;
    \vspace{2pt}
    \!update $x_i^*(k\!+\!h|k)$, \!$u_i^*(k\!+\!h|k)$ at future hubs\;
    \vspace{1pt}
    \!\textbf{return} $\textbf{\textit{u}}_i^*(k)$, $\{\mathcal{R}_i^*(k\!+\!h|k)\}$.
    }
  }
  \caption{Event-triggered Distributed MPC for Platoon Coordination at Hubs\vspace{2pt}}\label{Alg.1}
\end{algorithm}

\begin{myrem}
With the knowledge of predictions $\hat{\textbf{\textit{x}}}_{-i}(H_i(k\!+\!h|k))$ and $\hat{\textbf{\textit{u}}}_{-i}(H_i(k\!+\!h|k))$ of truck $i$'s related trucks, the predicted partner set $\mathcal{R}_i(k\!+\!h|k)$ of truck $i$ is a function of $\textbf{\textit{u}}_i(k)$ by (\ref{equ.8}). Therefore, in \textbf{Algorithm \ref{Alg.1}}, the optimal predicted partner sets corresponding to the optimal $\textit{\textbf{u}}_i^*(k)$ are denoted by $\{\mathcal{R}^*_i(k\!+\!h|k)\}$ with $h\!\in\![0\!:\!N_i\!-\!2\!-\!k]$.
\end{myrem}
\vspace{2pt}

%%%%%%%%%%%%%%%%%%%%%%%%%%%%%%%%%%%%%%%%%%%%%%%%%%%%%%
%--------Section IV. Numerical Example
\section{Simulation over Swedish road network}
\begin{figure}[thpb]
     \centering
     \includegraphics[width=0.42\linewidth]{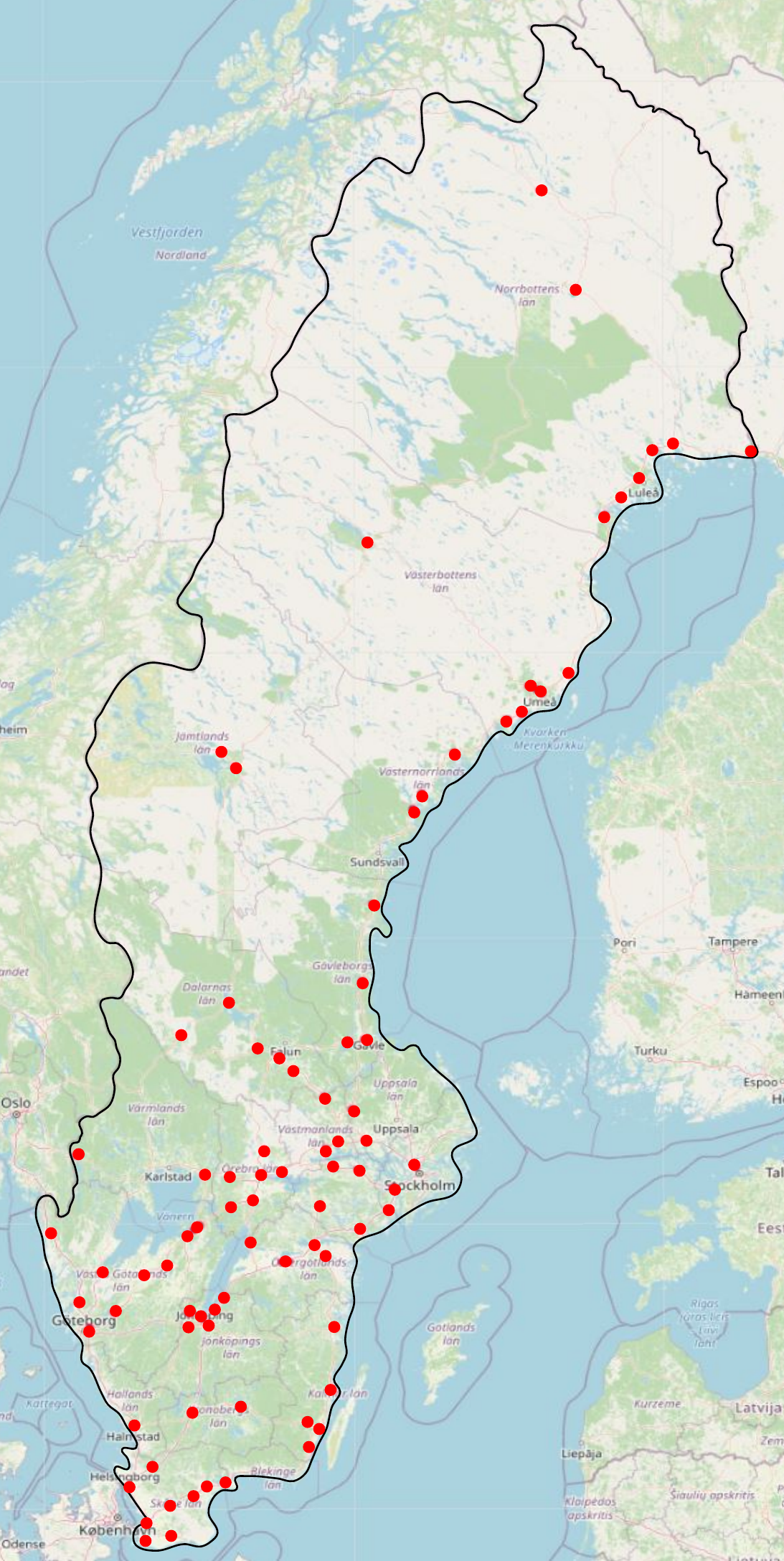}
     \vspace{-2pt}
      \caption{The considered hubs are marked out on the map of Sweden.}
      \label{Fig.4}
   \end{figure}

This section provides a simulation over the Swedish road network to demonstrate the effectiveness of the proposed algorithm. The number of trucks in the simulation is $100$, and for each of the trucks we randomly select its origin and destination from the $84$ major hubs in the road network of Sweden. The considered hubs are shown by the red nodes in Fig.~\ref{Fig.4}. The routes and travel times of individual trucks are obtained from  \textit{OpenStreetMap}~\cite{OpenStreetMap}. 
\vspace{-1pt}
   
We assume that each truck starts its trip at a random time between 8:00  a.m. and 9:00 a.m. The total maximal waiting time of the trucks along their entire trips is $1$ hour, while the maximal waiting time of each truck at a local hub is $30$ minutes. Since the origins and destinations of individual trucks are generated without restrictions on the total travel times, to avoid fatigue driving, we assume that drivers will be changed at hubs if the experienced travel times violate the maximized driving time restrictions.
\vspace{1pt}

The platooning benefit is assumed to be the reduced fuel consumption of follower trucks, and we assume that the fuel consumption of follower trucks is reduced by $10~\%$. This will lead to a monetary saving of $0.72$ Swedish kronor (SEK) per follower per kilometer. The velocity of trucks on motorways is considered as $80$ kilometers per hour, and thereby, the platooning benefit is $\xi_i\!=\!57.6$ SEK per hour. On the other hand, the cost of waiting at hubs is assumed to be low since the delivery deadlines are respected. In this example, we set the cost of waiting as  $\epsilon_i\!=\!45$ SEK per hour.
\vspace{2pt}

\begin{figure}[thpb]
     \centering
     \includegraphics[width=0.97\linewidth]{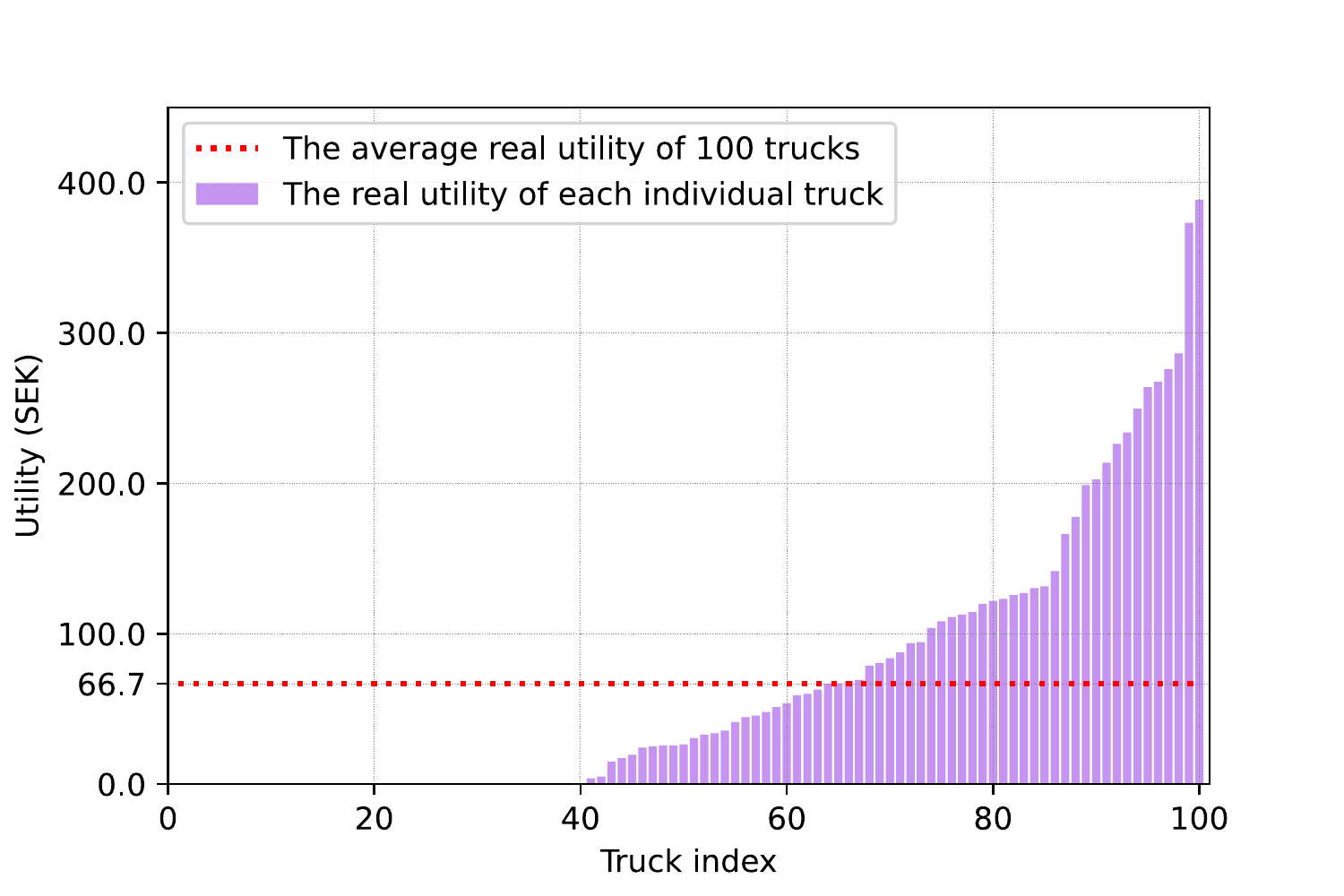}
      \vspace{-5pt}
      \caption{The real utilities of individual trucks}
      \label{Fig.5}
\end{figure}

\begin{figure}[thpb]
     \centering
     \includegraphics[width=0.97\linewidth]{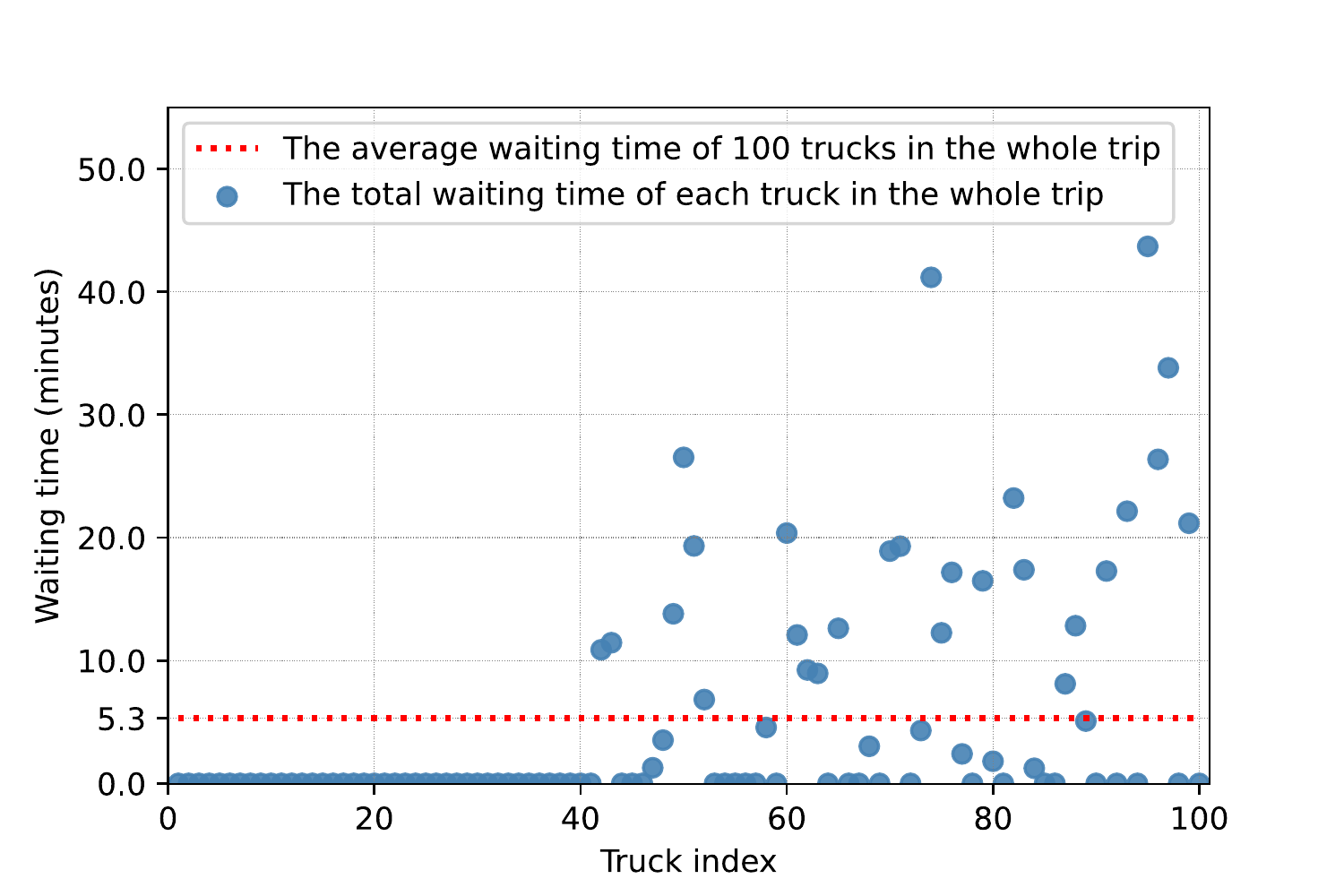}
      \vspace{-5pt}
      \caption{The total waiting times of individual trucks at hubs.}
      \label{Fig.6}
\end{figure}

\begin{figure}[H]
     \centering
     \includegraphics[width=0.97\linewidth]{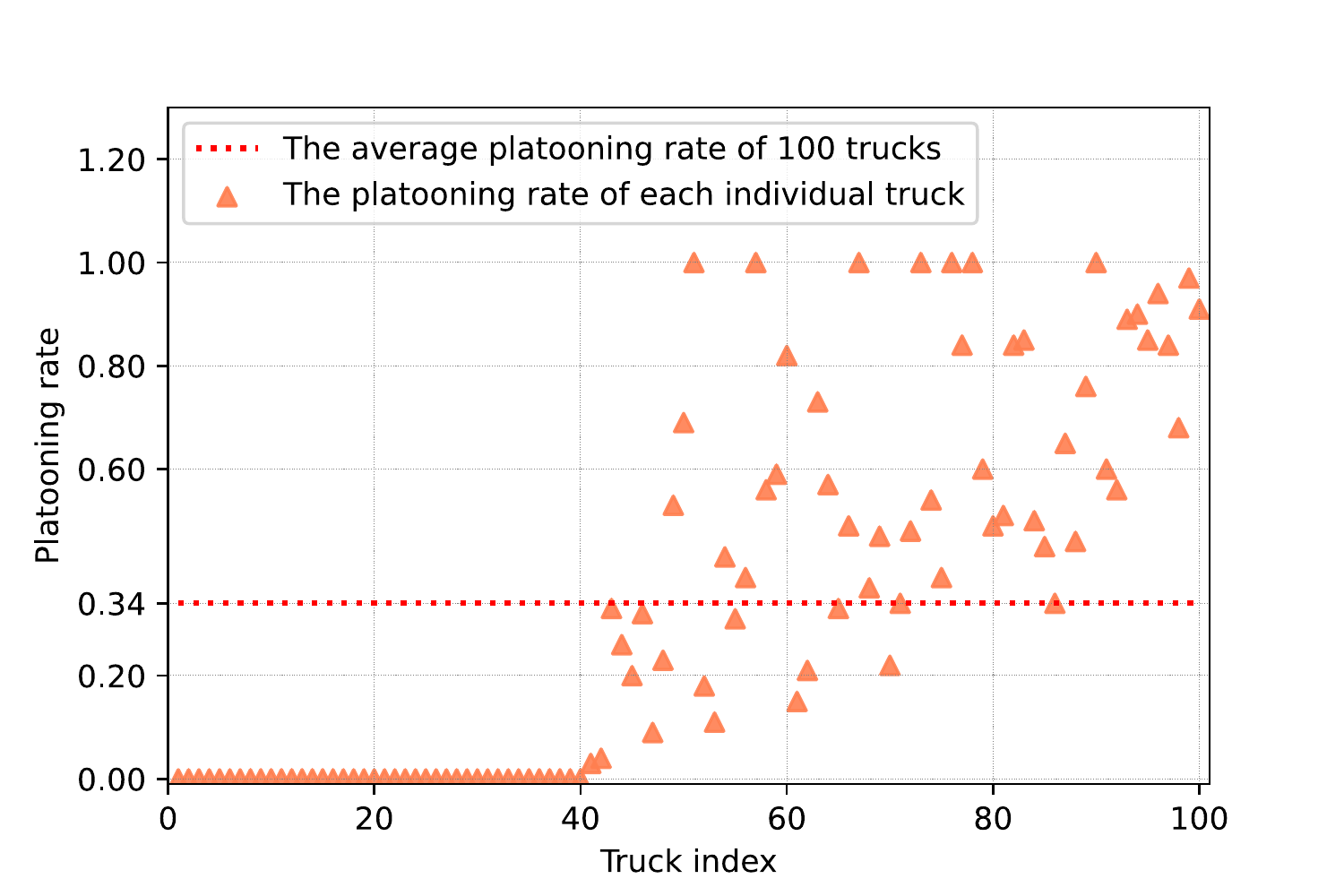}
     \vspace{-5pt}
      \caption{The platooning rates of individual trucks.}
      \label{Fig.7}
 \end{figure}

\begin{figure}[H]
     \centering
     \includegraphics[width=0.97\linewidth]{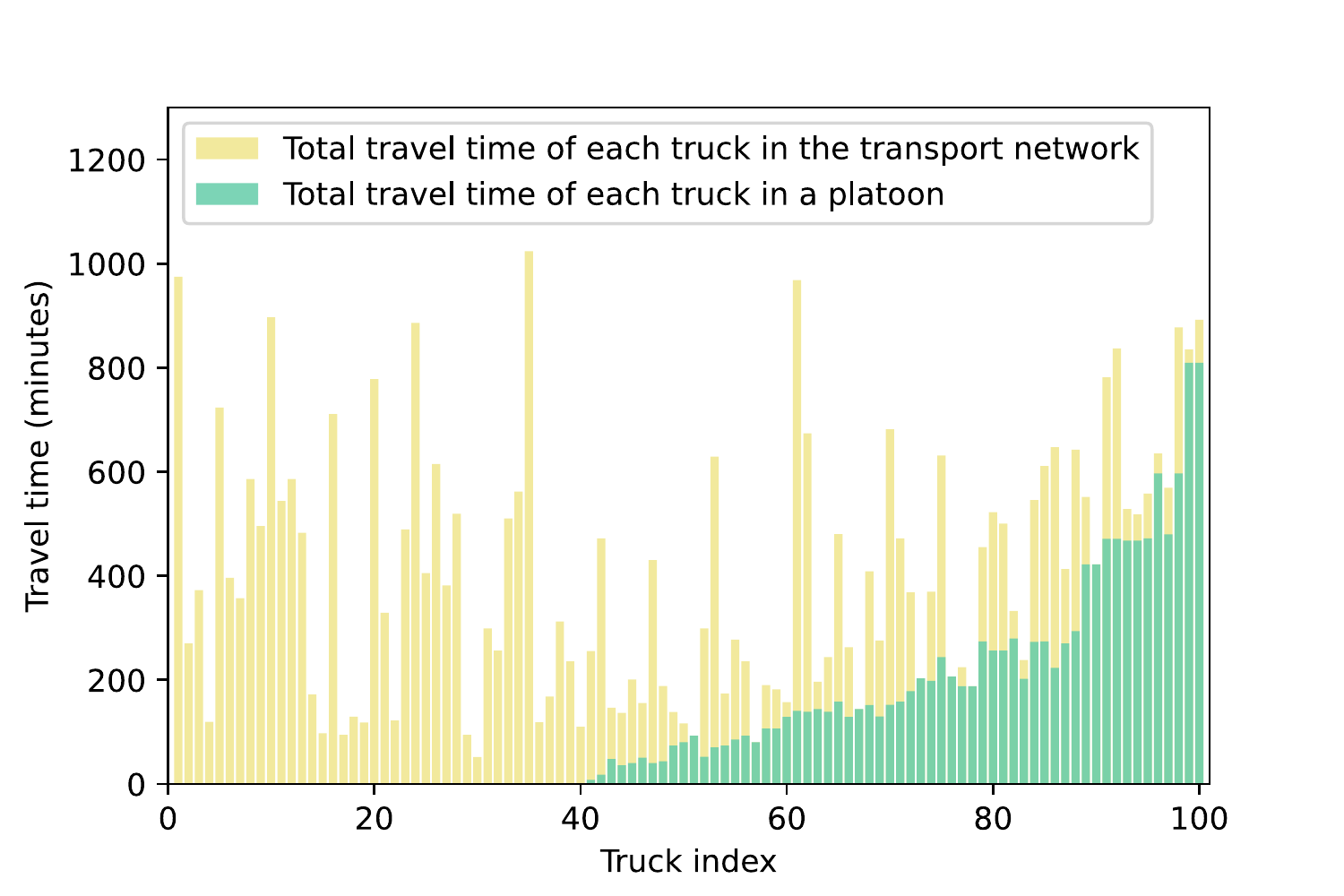}
     \vspace{-5pt}
      \caption{The total travel times of individual trucks.}
      \label{Fig.8}
 \end{figure}

\begin{figure}[H]
     \centering
     \includegraphics[width=0.97\linewidth]{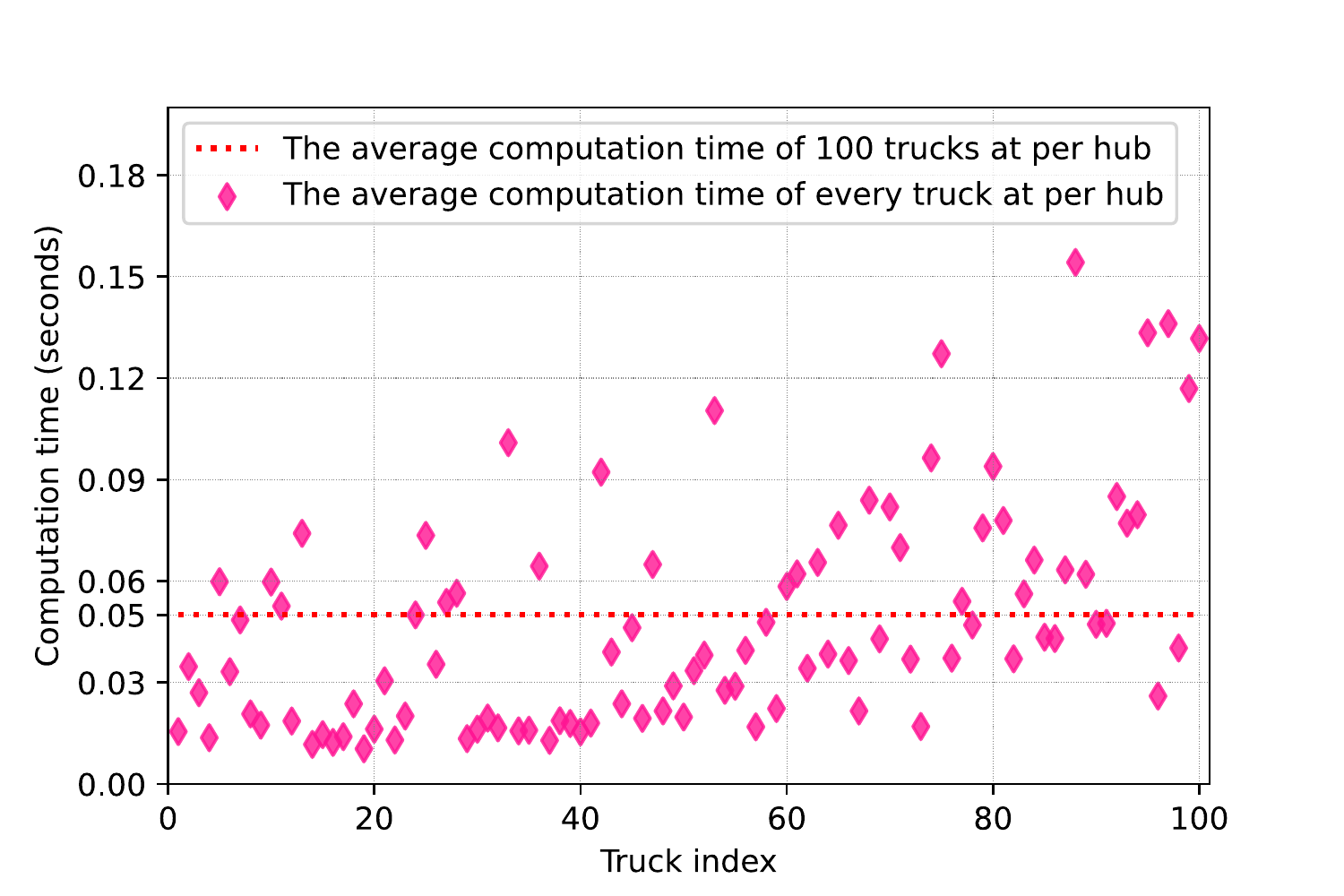}
     \vspace{-5pt}
      \caption{The computation time of individual trucks.}
      \label{Fig.9}
\end{figure}

The simulation results of every individual trucks are provided in Fig.~\ref{Fig.5}--\ref{Fig.9}. Fig.~\ref{Fig.5} shows the realized total utility of each of the trucks and the truck indexes are sorted according to trucks' utilities. The figure shows that the utilities range between $0$-$400$ SEK, and that approximately $60 \%$ of the trucks have a non-zero utility even though there were only $100$ trucks in the transport network. 

The waiting times of the trucks are given in Fig.~\ref{Fig.6}. It is seen in the figure that the trucks with zero utility do not wait, and the waiting times of the rest of the trucks are fairly uncorrelated with their utilities. The average waiting time of the trucks in their whole trips is $5.3$ minutes.

Subsequently, to evaluate the platoon coordination efficiency, each truck's time as alone driving is compared with its time as a platoon member, and we define the platooning rate for each truck $i$ as 
\begin{equation}
     \textbf{\textit{r}}_i=\frac{\text{Total travel time of truck $i$ in a platoon}}{\text{Total travel time of truck $i$ in the network}}.\label{equ.18}
 \end{equation}

The platooning rates of every individual trucks are shown in Fig.~\ref{Fig.7}, from which we can see the average platooning rate of the $100$ trucks is $0.34$. In addition, around $45 \%$ of the trucks have a higher platooning rate than the average value. The total travel times in the network and the times in a platoon of the trucks are shown in Fig.~\ref{Fig.8}.
 
Fig.~\ref{Fig.9} shows the average computation time of each truck used to solve the distributed MPC problem at the hubs. As is shown, the average computation time of the $100$ trucks at hubs is $0.05$ seconds, which indicates a high efficiency of the proposed algorithm. Moreover, Fig.~\ref{Fig.7} and Fig.~\ref{Fig.9} indicate that the computational time slightly correlates with the platooning rate of a truck. The above simulation results together demonstrate the effectiveness and characteristics of the developed method.

%%%%%%%%%%%%%%%%%%%%%%%%%%%%%%%%%%%%%%%%%%%%%%%%%%%%%%%%%%%%%%%%%%%%%%%%%%%%%%%
%-----------Section V. Conclusions and Future Works
\section{Conclusions and Future Work}
In this paper, an event-triggered distributed MPC approach was developed to address the hub-based platoon coordination in large-scale transport systems. The scheduling of waiting times at hubs was formulated as a distributed MPC problem. The utility of individual trucks consists of the predicted reward from platooning and the predicted loss caused by waiting at hubs. In this distributed framework, every truck is able to make decisions independently according to limited information and optimize its waiting times at different hubs. The proposed distributed MPC method has the merits of light computing load and is suitable for the optimization of massive networks. Finally, the numerical example of one hundred trucks in Swedish transport system verified the effectiveness of the approach.
\vspace{1pt}

The distributed MPC framework for platoon coordination carried out in this paper can be further extended in different ways. One of the interesting directions will be the distributed cooperative platoon for trucks from multiple fleets, and another possible extension could be taking into account stochastic travel times in the platoon coordination.
\vspace{5pt}

%%%%%%%%%%%%%%%%%%%%%%%%%%%%%%%%%%%%%%%%%%%%%%%%%%%%%%%%%%%%%%%%%%%%%%%%%%%%%%%
%-----------Section VI. Acknowledgment
\section{Acknowledgment}
The authors would like to thank Yuchao Li for the helpful discussions, the anonymous reviewers for their insightful comments, especially on the reward function, and the support of the career development scholarship from Shanghai Jiao Tong University.
%%%%%%%%%%%%%%%%%%%%%%%%%%%%%%%%%%%%
\bibliographystyle{IEEEtran}
\bibliography{Ref}
\end{document}